\documentstyle[aps,prl,psfig,epsfig]{revtex}
\newcommand{\be}{\begin{equation}}
\newcommand{\ee}{\end{equation}}
\newcommand{\bea}{\begin{eqnarray}}
\newcommand{\eea}{\end{eqnarray}}

\begin{document}
\draft
\title{ Dynamics of a tunneling magnetic impurity: Kondo effect induced 
incoherence}
\author{L. Borda,$^1$ G. Zar\'and,$^{1,2}$} 
\address{
$^1$Research Group of the Hungarian Academy of Sciences, Institute of Physics, 
TU Budapest, H-1521 Hungary, 
\\
$^2$Lyman Physics laboratory, Harvard University, Cambridge MA 02145
}
\twocolumn[\hsize\textwidth\columnwidth\hsize\csname
@twocolumnfalse\endcsname

\date{\today}

\maketitle

\begin{abstract}
We study how the formation of the Kondo compensation cloud 
influences the dynamical  properties of a magnetic impurity 
that  tunnels  between two positions in a metal. The Kondo effect 
dynamically generates a strong tunneling impurity-conduction electron  
coupling, changes the temperature dependence of the tunneling rate, 
 and may ultimately result in the destruction of the  coherent motion 
of the particle at zero temperature. We find an interesting  
two-channel Kondo fixed  point as well for a vanishing overlap between the 
electronic states that screen the magnetic impurity. We propose a
number of  systems where the predicted features  could be observed.
\end{abstract}
]
\narrowtext

Tunneling of a heavy particle or some collective degree of freedom in a 
dissipative environment has been the subject of intense 
theoretical and experimental research in the past 
and is by now  reasonably 
well-understood 
\cite{Leggett,Kagan,book,saleur,costizarand,giordano,golding}. 
Maybe the most intriguing  case is that of  {\em  ohmic} dissipation, 
where the particle usually  couples to  electron-hole excitations of 
a metallic environment.  In this case the bare tunneling amplitude $\Delta_0$ 
of the particle is  strongly  renormalized 
 due to the dissipative environment,  and becomes temperature dependent. 
In the simplest scenario, where the tunneling 
occurs between two sites, the effective tunneling 
 displays a power-law behavior over a wide range of temperatures,  
$\Delta(T) \sim \Delta_0  (T/\omega_0)^{\alpha}$, 
with $\omega_0$ a  high-energy cut-off of the 
order of the Debye frequency, and $\alpha$ a  
dimensionless coupling  constant \cite{Leggett,Kagan}. 
The renormalization of the tunneling  amplitude is a  consequence 
of Anderson's orthogonality  catastrophe\cite{Anderson}:  At any position 
the presence of the particle generates a screening cloud 
that consists of an  {\em infinite} number of electron-hole 
excitations. The formation of this   huge 
'electronic polaron-cloud' slows down the particle, increases 
its mass, and thus decreases its tunneling amplitude. 
Depending on the specific value of the coupling  $\alpha$,
the dynamics of the particle can be of three different  kind: 
(a) If the coupling is small the particle moves with slightly damped
{\em coherent} oscillations between  the two positions at $T=0$. 
(b) For $1/2 < \alpha < 1$ the motion of the particle becomes 
{\em incoherent}, while for even larger couplings (c) 
the particle becomes localized and cannot move from one well 
to the other. 

At low temperatures and larger values of  $\alpha$, 
more complicated  dynamical processes involving simultaneous 
tunneling and  electron scattering (electron assisted 
tunneling) may become more relevant and can, in principle, delocalize 
the particle or lead to  other types of strongly correlated 
states \cite{Zawa,Fisher}. In this paper, however, we focus on 
the rather generic case of 'slow' TLS's \cite{Zawa}, 
where the energy scale generated by these processes is extremely small, 
and they can therefore be neglected. 

In the present paper we focus our attention to the 
very interesting but  poorly understood case of 
a tunneling {\em magnetic impurity} coupled to an ohmic 
environment \cite{zar,ye96}. Possible examples of such a system 
include a magnetic impurity  tunneling between an STM tip and a 
metallic  surface\cite{sethna},  a Kondo impurity in an amorphous 
region\cite{ralph}, a spin 1/2 quantum soliton interacting 
with a metallic environment\cite{polyac}, 
or charge tunneling in double  quantum-dot systems\cite{future,sakai}.   
In all these cases the spin of a magnetic  impurity couples to the conduction 
electrons through an exchange interaction. This exchange interaction  becomes 
renormalized as the temperature is lowered  and may lead to the dynamical 
formation  of a Kondo compensation cloud. Since the tunneling particle 
has to drag this cloud with itself, the exchange coupling will 
lead to a dynamically generated temperature dependent renormalization of 
the tunneling amplitude. In the present paper we study 
this interplay between the Kondo effect and the orbital motion of the TLS
in detail.

For the sake of simplicity  we focus our attention to 
the simplest possible case of a TLS, where 
tunneling  takes place  between two positions only, 
${\bf R}_{\pm}$.  Furthermore, though we also  discuss 
the  role of asymmetry to some extent,  
we mostly focus on spatially  symmetrical TLS's. 
We show, in particular, that the Kondo effect associated 
with  the magnetic degrees of freedom leads to a strong 
{\em temperature-dependence} of  the exponent $\alpha$, and may 
eventually induce a decoherent  state. 
Destroying the Kondo cloud with a  magnetic field one could 
then be able to drive the particle back to the coherent regime. 
The orbital motion, on the other hand,  may  lead to the 
appearance of a {\em two-channel Kondo} state under 
special conditions, where the impurity tunnels very fast back and forth 
and forms a Kondo state with the conduction  electrons at  {\em both} 
positions.

 We describe the TLS by the tunneling Hamiltonian 
\begin{equation} 
H_{\rm tun} = -(\Delta_0 T^x + \Delta_z T^z) \;,
\end{equation}
where the two pseudospin states  $T^z = \pm 1/2$
correspond to the two tunneling positions, $\Delta_0$ is the tunneling 
matrix element, and $\Delta_z$ describes the asymmetry of the TLS. 

As a first approximation, we  consider the exchange  interaction between  
the tunneling particle   and the electrons as completely  local: 
 \begin{eqnarray}
H_{\rm int} &=& \sum_{q =\pm}
J_q \;P_q\; {\vec S}\; (\Psi^{\dagger}_q {\vec \sigma}\Psi_q)\;.
\label{eq:hint}
\end{eqnarray}
Here $P_\pm=1/2 (1 \pm 2T^z)$  projects out the TLS  states 
at positions ${\bf R}_{\pm}$,  $J_\pm$ is the exchange coupling at 
these positions, ${\vec S}$ denotes the spin operator of the impurity, and 
$\vec \sigma$ stands for the Pauli matrices. 
The field operators,  $\Psi^\dagger_{\pm,\mu}=\int 
e^{i{\bf kR}_\pm} c^\dagger_{{\bf k}\mu} {\bf d}^3{\bf k} /{(2\pi)^3}$, 
create conduction electrons  at ${\bf R}_\pm$ with  spin $\mu$.  
Including more complicated non-local interactions in Eq.~(\ref{eq:hint})
such as spin-assisted tunneling  has not changed  our results 
\cite{zar,future}. 

First we consider the case of a  symmetric TLS with 
$J_-=J_+=J$ and $\Delta_z=0$.  Following the standard procedure 
\cite{Fisher,libero}, the relevant  conduction  electron degrees 
of freedom can be  represented simply by one-dimensional 
fermion fields \cite{Kagan}, $c_{p\alpha\mu}$,  
obeying canonical anticommutation  relations,   
$\{c_{p \alpha\mu}, c^\dagger_{p' \alpha'\mu'}\}  = 2\pi \;\delta(p-p')  
\delta_{\alpha\alpha'} \delta_{\mu\mu'}$, and
described by  
\begin{equation}
 H_{\rm el} =   
\sum\limits_{\alpha=e,o}
\sum\limits_{\mu =\uparrow,\downarrow} 
\int_{-k_F}^{k_F} {dp\over {2\pi}}\; 
v_F\; p\; c^{\dagger}_{p\alpha\mu}c_{p\alpha\mu}\;.
\label{eq:dirac}
\end{equation}
In Eq.~(\ref{eq:dirac}) the  radial momentum  $p$ is measured 
from  the Fermi momentum $k_F$, $v_F$ is the Fermi velocity, 
 $\alpha =\{ e,o\}$ is the parity,   and $\mu$ denotes the spin.  
Introducing the  fields   $\Psi_{\alpha,\mu} \equiv   
\int_{-k_F}^{k_F}   \; c_{p\;\alpha\;\mu} {dp/{\sqrt{2\pi}}}$,   
Eq.~(\ref{eq:hint}) becomes
\begin{eqnarray}
H_{\rm int} &=& {g\over 2} \vec S(1+F)\Psi^{\dagger}_{e} {\vec \sigma}
\Psi_{e} +{g\over 2} \vec S(1-F)\Psi^{\dagger}_{o}
{\vec \sigma} \Psi_{o}\nonumber\\
&+&g{\vec S}\sqrt{1-F^2}\; {T}^z
\left(\Psi^{\dagger}_{e}
{\vec \sigma}
\Psi_{o}+\Psi^{\dagger}_{o}
{\vec \sigma}
\Psi_{e}
\right)\;.
\label{eq:H_int_eo}
\end{eqnarray}
Here  $F= \sin (k_F d) / {k_F d}\;$  measures the overlap  
of the states $\Psi_{\pm}$,  with $d =  |{\bf R}_+-{\bf R}_-|$ the 
tunneling distance, and $g = J k_F^2/2\pi^2$.  For $d=0$ this Hamiltonian 
obviously reduces to  the single-channel  Kondo model. 
In the following we set $v_F = k_F = \hbar  = 1$. 

To study the zero temperature dynamical 
properties  of the tunneling Kondo spin we used Wilson's
numerical renormalization group (NRG)~\cite{nrg}. 
In this very accurate numerical technique one constructs  
a series of Hamiltonians, $H_N$, which are diagonalized 
iteratively. Having obtained  the many-body eigenstates
and energies  of $H_N$ one can use them to
calculate physical quantities at an energy 
scale $T,\omega\sim\omega_N\sim \Lambda^{-(N+1)/2}$, 
with $\Lambda \approx 3$ a discretization parameter.
Our results were obtained by keeping the lowest
$250$ states in each iteration. To obtain accurate results we 
exploited the following  symmetries: (i) parity  (ii) global  
spin rotations, and   (iii) a hidden $SU(2)$ symmetry,
related to   electron-hole symmetry \cite{Jones}. 

We computed the $T=0$ impurity spin and pseudospin spectral functions, 
$\varrho_{\cal O}^i = - {\rm Im} \{\chi_{\cal O}^i(\omega)\}$, 
(with $\chi_{\cal O}^i(\omega)$ the Fourier transform of the retarded 
response function), using their  Lehmann representations: 
 \begin{eqnarray}
\varrho_{\cal O}^i(\omega>0) &=&\sum\limits_n | \langle n |{\cal O}^i| 
0\rangle|^2 \delta (E_n-E_0-\omega).\nonumber
\end{eqnarray}
Here  $\mid 0\rangle$ ($\mid n\rangle$) is the ground 
($n$th excited) state of the 
system with energy $E_0$ ($E_n$), and  ${\cal O}^i$ denotes the operators 
$T^i$ and $S^i$. 
\begin{figure}[tb]
\begin{center}
\epsfxsize6.5cm
\epsfbox{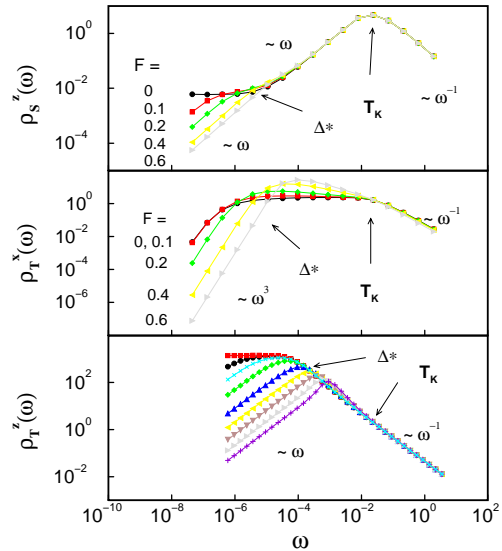}
\end{center}
\caption{\label{fig:log} 
Logarithmic plot of  various spectral functions discussed in the 
text for $\Lambda = 3$, 
$g = 0.144$, and  $\Delta_0 = 2.31\cdot 10^{-5}$. 
The energy scales $T_K$ and  $\Delta$ are also indicated.
}
\end{figure}

Our results are summarized in Fig.~\ref{fig:log}.  Let us first 
focus on the somewhat peculiar case of  $F=0$, 
where $\Psi_+$ and $\Psi_-$ do not overlap.  
In this case we  can observe two distinct cross-overs:
The first  takes  place at the Kondo energy  $T_K\approx e^{-1/2g}$ 
and corresponds to the formation of a  local Kondo 
state at the {\em actual} position of the TLS. 
Above $T_K$ all spectral functions behave as $1/\omega$, 
indicating that all correlation functions are constant 
for times shorter than $1/T_K$.
Below $T_K$ the spin spectral function becomes linear. Performing a 
Hilbert transform on can see that this   corresponds to a 
constant impurity susceptibility $\sim 1/T_K$.  
The $z$-component of the pseudospin spectral function is practically
unaffected by the formation of the Kondo compensation cloud: 
at this time scale  tunneling events are very rare, and the 
particle  can be considered as immobile. This interpretation is confirmed 
by a detailed analysis of the finite size spectrum\cite{future}. 
The logarithmic slope of the  $x$-component of the pseudospin 
spectral function,  however, does become  modified:  
this change is related to the dynamical renormalization of the 
tunneling  amplitude by the formation of the Kondo compensation cloud.

Though  $\Delta_0\ll T_K$ in our computations, 
 $\Delta_0$ is a relevant perturbation,  and leads to a second 
cross-over at a  renormalized value of the tunneling amplitude, 
$\Delta^*$,  where the TLS freezes into the even tunneling state. 
For $F = 0$ a two-channel Kondo state is formed below 
$\Delta^*$, as confirmed by the analysis of the 
finite size spectrum \cite{future}. 
This is most easily understood by observing that 
the last term of Eq.~(\ref{eq:H_int_eo}) flips the TLS between
the $T^x = \pm1/2$ states, and can therefore be dropped 
below $\Delta^*$.  Here $H_{\rm int}$ becomes simply the 
two-channel Kondo Hamiltonian, and a two-channel Kondo state is 
formed in the spin sector \cite{Zawa}. The spin spectral function and 
the $z$-component of the pseudospin  spectral function
become constant below $\Delta^*$, implying the logarithmic 
divergence of the spin- and $T^z$-susceptibility at low 
temperatures, $\chi_S(T) \sim \Delta^* \ln(T_K/T) /T_K^2$, and 
$\chi_T^z(T) \sim \ln(T_K/T)/ \Delta^*$, and the asymptotic 
behavior  of the time-dependent correlation functions 
$\langle  S^i(t) S^i(0)\rangle   \sim\langle T^z(t) T^z(0)\rangle \sim 1/t$.

These results immediately imply that the external magnetic field and 
asymmetry are both relevant operators  of scaling dimension $1/2$ 
at this two-channel Kondo  fixed point\cite{cardy}. 
The two-channel Kondo state is thus extremely unstable. 
It  crosses over to a Fermi liquid state also for any finite overlap 
$F$: For small values of $F$  a third  
crossover occurs at an energy $T^*$ well below $\Delta^*$, however, 
for generic $F$'s this second crossover takes place almost 
simultaneously with the crossover at $\Delta^*$ and only a small kink 
remains from the two-channel Kondo behavior of $F=0$.

\begin{figure}
\begin{center}
\epsfxsize6cm
\epsfbox{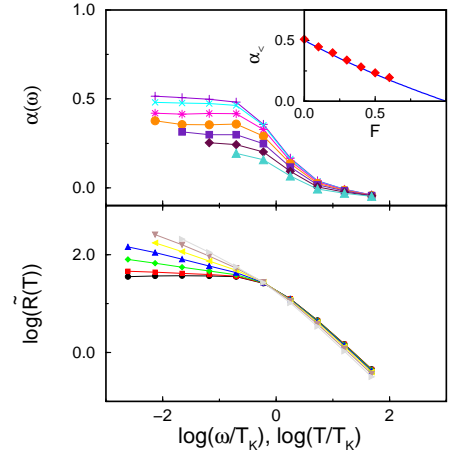}
\end{center}
\caption{\label{fig:alpha} 
Top: Energy- and overlap-dependence of the anomalous dimension $\alpha$ of the 
dimensionless tunneling amplitude for $F=0$, 0.1, 0.2 0.3, 0.4, 
0.5, and 0.6 (top to bottom),  as determined from the 
logarithmic derivative of the spectral function 
$\varrho_\tau^x$. Bottom: Temperature dependence of the normalized 
tunneling rate, $\tilde R (T) = R(T)/\Delta_0^2$. $F$ decreases 
from top to bottom.   
We used  $\Lambda = 3$, $g  = 0.144$, and  $\Delta_0 = 2.31\cdot 10^{-5}$ 
in both figures.
}
\end{figure}

To determine the scale $\Delta^*$ and the 
temperature dependence of the tunneling 
rate we invoked  renormalization group arguments. 
The Hamiltonian involves three dimensionless parameters:
$g$, $F$, and $\Delta_0/\omega_0$, with $\omega_0$  the 
high-energy cutoff.  
Well above $\Delta^*$,  $\Delta_0/\omega_0$ is small and 
the effective tunneling at energy scale $\omega$
or temperature $T\sim \omega$, $\Delta(\omega)$, 
satisfies the  following  scaling equation:  
\begin{equation} 
{d\ln \Delta(\omega)\over d\ln(\omega_0/\omega)} = 
- \alpha(g(\omega),F)\;,\;\;\;(\omega_0 \gg \Delta(\omega)).
\label{eq:scaling}
\end{equation}
Above $\Delta^*$ the impurity is immobile.
Therefore the effective Kondo coupling $g(\omega)$ 
does  not depend   on the other two parameters, and  
is a    universal function of  $\omega/T_K$.  In  this regime  $F$
can be completely transformed out of the Hamiltonian  
and is therefore constant. 
Thus  $\alpha = 
\alpha(\omega/T_K,F)$ is a universal  function of $\omega/T_K$
and $F$ for $\omega_0 \gg \Delta $. This function   is related to 
the time-dependent   correlation  function as 
$\langle T^x(t) T^x(0)\rangle \sim  1/t^{2\alpha(\omega)}$ 
($t \sim 1/\omega$) \cite{cardy}.  Assuming that the retarded and  
the time ordered correlation functions have the same singular 
behavior, this immediately implies that, within logarithmic accuracy,  
$\alpha$ can be determined from  the  logarithmic  
derivative of $\varrho_T^x$ as
\begin{equation}
\alpha(\omega) \approx {1\over 2} 
\bigl( {d\ln\;\varrho_T^x \over d\ln \omega} + 1\bigr)\;, \;\;\; 
(\omega \gg \Delta^*).
\label{eq:alpha}
\end{equation}
We plotted this universal function in Fig.~\ref{fig:alpha}. 
For large frequencies  $\alpha\approx  0$, meaning that the tunneling 
amplitude  remains unrenormalized above $T_K$. Below $T_K$, 
on the other hand, it scales to an  overlap-dependent 
constant, $\alpha_<(F)$.  As shown in the inset,  
$\alpha_<$ coincides 
with the  Anderson  orthogonality  exponent $K$ for a
maximally strong scatterer with a phase shift 
$\delta_\uparrow = \delta_\downarrow = \pi/2$ \cite{yoshida}, 
\begin{equation}
\alpha_< \equiv  
\sum_\sigma\;
\Bigl({1\over \pi}\; {\rm atan}\;{{\rm tan}\;\delta_\sigma \;\sqrt{1 - F^2} 
\over \sqrt{1 + F^2\; {\rm tan}^2 \delta_\sigma }}\Bigr)^2\;.
\label{eq:yoshida}
\end{equation}
Thus, far below $T_K$ (but still  above $\Delta^*$) the  Kondo 
impurity acts as a maximally  strong potential scatterer 
in agreement with Nozi\`eres' Fermi liquid  picture \cite{Nozieres}. 

A  quantity of primary interest is
the renormalized temperature dependent tunneling 
rate,  which can be experimentally 
determined  by performing real-time  measurements~\cite{golding}.  
By dimensional analysis, for an ohmic heat bath 
it is given by  $R(T) \sim  \Delta(\omega = T)^2/T$ \cite{Leggett}. 
We computed  it by integrating  Eq.~(\ref{eq:scaling}) 
(Fig.~\ref{fig:alpha}, bottom).  
As the most  striking consequence of  the Kondo effect,   
the logarithmic slope of $R(T)$ changes at $T\approx T_K$.

\begin{figure}
\begin{center}
\epsfxsize5.8cm
\epsfbox{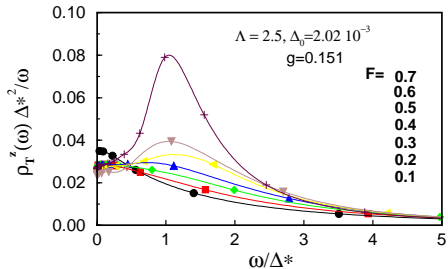}
\end{center}
\caption{\label{fig:coherence} 
Rescaled spectral functions and the evolution of the  coherence peak 
with increasing overlap. We computed $\Delta^*$ from 
Eq.~(\protect{\ref{eq:delta*}}) with $C=0.0916$ and $T_K = e^{-1/2g}$.
}
\end{figure}

The energy scale  $\Delta^*$ is determined \cite{Leggett}
by the condition   $\Delta(\omega=\Delta^*)  \approx  \Delta^*$,
leading to
\begin{equation}
\Delta^* = 
\Delta_0 \Bigl(C {\Delta_0\over T_K}\Bigr)^{\alpha_</(1-\alpha_<)}\;,
\label{eq:delta*}
\end{equation}
with $C$ a constant of the order of unity. 
The constant $\alpha_<$ also characterizes the dissipative 
non-equilibrium  dynamics of the  TLS below $T_K$ \cite{Leggett}.  
In Fig.~\ref{fig:coherence} we show the rescaled  spectral 
function, ${\Delta^*}^2 \varrho_T^z(\omega)/\omega$, related
to the real part of the retarded response function.  
Without dissipation,  $\alpha_<=0$ ($F=1$), the  tunneling of the TLS 
is entirely coherent: the TLS oscillates between the two positions
without damping, and the spectral function consists of two Dirac delta's.
For $F<1$ the coherence peak broadens:  The oscillations  become 
exponentially damped and  at very  long time scales (at $T=0$) the 
correlation  function behaves as $\langle T^z(t) T^z(0) 
\rangle \sim 1/t^2$. For even smaller values of overlap the  
peak becomes completely  invisible, and the formation of the Kondo 
compensation cloud  suppresses the coherent motion.   

In the general case, the TLS model is more complex than 
the one we discussed until now. The TLS is not symmetrical: 
$\Delta_z \ne 0$,  and the couplings $J_\pm$ in Eq.~(\ref{eq:hint})  
are not equal.   The difference between $J_\pm$ leads to {\em two } 
subsequent Kondo effects at $T_K$'s associated with  the two positions, and 
consequently  changes in the logarithmic slope  of $R(T)$ twice, 
while a finite  $\Delta_z$ generates a new energy scale, below 
which the TLS freezes into one of the states
$T_z=\pm1/2$ \cite{future}. 

In Eq.~(\ref{eq:hint}) we only took into account the 
exchange interaction. In reality, the TLS and 
the electrons  interact through a local potential scattering as well. 
However, for a Kondo impurity this potential scattering
is relatively small (of the order of the exchange  
coupling) in the  exchange scattering-channel
 and since it remains  essentially unrenormalized, 
its effect  can be neglected compared to that 
of the exchange  interaction. 
Potential scattering  in {\em other} scattering channels may, however,
 be still  present and  shift $\alpha$ by a temperature-independent value 
$\alpha(T)\to \alpha(T) + \alpha_>$. Therefore, for small 
overlaps one can easily be in a situation, where 
the Kondo effect drives the TLS from a coherent ground
state to an incoherent state with $\alpha >1/2$ (see Fig.~\ref{fig:alpha}). 

For small enough Kondo temperatures, the value of 
$\delta_\sigma$ in Eq.~(\ref{eq:yoshida})  can be changed by an  
external magnetic field \cite{Nozieres}, which gradually 
destroys the Kondo effect, thereby  changes the logarithmic slope of 
$R(T)$ to its value above $T_K$, and restores the  
coherent TLS motion \cite{future}. 

In the experimental realizations mentioned  in the introduction 
it is the Kondo spin that moves and carries the compensation cloud. 
Our discussion can be, however, easily generalized to the case  
of a TLS that happens to be close to  a Kondo impurity 
\cite{Keijsers,Comm}.  In this case  Friedel  oscillations 
generated by the TLS  modify   the local  density of states at 
the Kondo impurity  and thus   the exchange  coupling will depend 
on the position of the TLS,  leading to a  Hamiltonian similar 
to Eq.~(\ref{eq:hint}). 

It is also an interesting question whether the two-channel Kondo 
behavior found can possibly be observed in realistic systems. 
Clearly,  one needs an almost perfectly  symmetrical TLS 
and a very small overlap $F$. In most  cases $F$ is typically large, 
excepting the magnetic impurity tunneling between an STM tip  and a 
surface,  where the electronic states coupled to the  spin hardly 
overlap at the two positions and $F\approx 0$. 
In order to observe  the two-channel Kondo behavior, 
one has to minimize the TLS asymmetry too:
The asymmetry $\Delta_z$ can be tuned to zero with the external 
voltage, while $J_+\approx J_-$ can be assured  
by using  sufficiently rounded tip  that is made  of the 
same  metal as the surface.

{\it Acknowledgements}:  
We are grateful to T. Costi, D.L. Cox, K. Damle, and E. Demler,  
for stimulating discussions, and especially  J.P. Sethna and B.I. 
Halperin, for  making useful suggestions.  
This research has been supported  by  
NSF Grants Nos. DMR-9985978 and DMR97-14725, and Hungarian Grants No. 
OTKA F030041,  T029813, and  T29236. 

\end{document}